# Producers of Popular Science Web Videos – Between New Professionalism and Old Gender Issues

Jesús Muñoz Morcillo[1]*, Klemens Czurda*, Andrea Geipel**,
Caroline Y. Robertson-von Trotha*

*ABSTRACT: This article provides an overview of the web video production context related to science communication, based on a quantitative analysis of 190 YouTube videos. The authors explore the main characteristics and ongoing strategies of producers, focusing on three topics: professionalism, producer's gender and age profile, and community building. In the discussion, the authors compare the quantitative results with recently published qualitative research on producers of popular science web videos. This complementary approach gives further evidence on the main characteristics of most popular science communicators on YouTube, it shows a new type of professionalism that surpasses the hitherto existing distinction between User Generated Content (UGC) and Professional Generated Content (PGC), raises gender issues, and questions the participatory culture of science communicators on YouTube.*

**Introduction**
Not very long ago YouTube was introduced as a platform for sharing videos without commodity logic. However, shortly after Google acquired YouTube in 2006, the free exchange of videos gradually shifted to an attention economy ruled by manifold and omnipresent advertising (cf. Jenkins, 2009: 120). YouTube has meanwhile become part of our everyday experience, of our "being in the world" (Merleau Ponty) with all our senses, as an active and constitutive dimension of our understanding of life, knowledge, and communication. However, because of the increasing exploitation of private data, some critical voices have arisen arguing against the production and distribution of free content and warning of the negative consequences for content quality and privacy (e.g., Keen, 2007; Welzer, 2016). Since online video consumption is one of the most widespread activities on the Internet, it is not surprising that the system also considers free video content as an economic factor in general. Competition between YouTubers for gaining public attention may have led to a new category of YouTube professional that we can no longer consider part of an amateur movement. However, what about popular science web videos? **[1]** Do competition, profit, and success also rule science communication on YouTube, or is science communication on YouTube an expression of democracy and participation beyond financial interests? In this sense, other cultural expectations are also challenged. Can we expect that online science communication on YouTube is not affected by typical gender issues? Also, what intentions lie behind the community-building measures of YouTube producers? To answer these questions, we must investigate the specific production context of science communicators on YouTube.

---

[1] jesus.morcillo@kit.edu (corresponding author)
* ZAK | Centre for Cultural and General Studies at the Karlsruhe Institute of Technology (KIT)
** Munich Center for Technology in Society – Technical University of Munich (TUM)

**State of the Art**
Researchers are currently beginning to analyze the characteristics of the video production context on YouTube. There are some studies on video interaction on online video networks (Benevenuto et al., 2009; Hanson et al., 2008), on video sharing (Cheng et al., 2007), as well as some works analyzing the participatory factors of YouTube (Burgess and Green, 2009), and the implications of the attention economics on this platform relating to information wars (Dauber, 2009). However, there are not many analyses that focus on the specific production context of popular science web videos. Insights into the production level can be gained through interviews with YouTube producers. In 2016 Erviti and Stengler conducted in-depth interviews with five professional producers behind the most important YouTube science communication channels from the UK. They revealed that the specific nature of online videos as an interactive science communication format is actively sought after by established TV producers and even pursued to a high level of performativity, building and feeding communities. Even if the advent of YouTube has made it possible for anyone to upload their own videos, the interviews conducted by Erviti and Stengler suggest that professional broadcasters like the BBC have the resources and the staff to exploit the real potential of science web videos (Erviti and Stengler, 2016: 12). However, the authors rightly state that as "inherent to an exploratory study with qualitative methods, no claim is being made regarding the generalization of these results" (Ervity and Stengler, 2016). Additional data via quantitative research is needed.

Erviti (2008) has recently published a content analysis on producers of online videos about climate change, vaccines, and nanotechnology by searching for these terms in the videos section of Google. She focuses on producer groups, including the type of producer, as well as age and gender of scientists, video formats, and objectives pursued by the producers. Despite the interesting approach, the author does not justify the different types of video producers identified, which are sometimes in overlapping categories, such as 'business' and 'television'. Also, a clear distinction or reassessment of the categories user generated content (UGC) and professionally generated content (PGC) is absent. The most interesting part of Erviti's contribution is the analysis of objectives, which builds on the work of Besley et al. (2016) on the identification of "nonknowledge objectives" such as entertainment for raising public awareness. Erviti's analysis confirms the identification of new subgenres in UGC made by Muñoz Morcillo et al. (2016). She reveals a possible interdependence between newsworthy topics and mass media, as well as between non-newsworthy topics and UGC or scientific institutions (Erviti, 2018: 37-38).

In addition to information about the producers' profile, the question arises as to who these producers are, in terms of gender and age. In its infancy, YouTube was presented as a democratic space, open to everyone, without the issues of traditional mass media. One of the democratic intents should be the equal representation of men and women. In a recent study on gender performativity on YouTube, Wotanis and McMillan (2014) state that women on YouTube are underrepresented (see also Molyneaux et al. 2008). The question is whether this could also be observed in science communication on YouTube, which a recent study by Amarasekara and Grant (2019) seems to confirm. Recently, Mike Thelwall and Amalia Mas-Bleda (2018) also analyzed the influence of a presenter's gender, and commenter sentiment towards males and females in 50 very popular channels, finding that some male commenters posted inappropriate sexual references that have the potential to alienate females. In previews research, Thelwall and his co-authors carried out statistical analysis on the behavior of scientists and commenters on TED videos (Sugimoto and Thelwall, 2013; Tsou et al., 2014) and on the role of online videos in research communication in general (Kousha et al., 2012).

Finally, combining interviews and ethnographic fieldwork, Geipel (2018) gained insights into the production process of five non-professional German YouTube science channels, questioning the influence of the platforms' politics. The author identified a platform-specific shaping of role models, production process, and communication logic in science communication. In particular, video producers, who do not belong to institutionalized science communication, are motivated by curiosity and entertainment and therefore create new formats.

**Purpose of this Paper**
Despite the previous research, the analysis of the YouTube production context is usually limited to particular issues. It focuses on a scientific topic, on a small amount of data, or it is limited to a qualitative level, where results cannot be generalized.

With this quantitative research, we present data from 190 YouTube videos about producers' professionalism, producer's gender and age profile, and community building. In addition, we compare the results against hypotheses derived from previous qualitative analysis and practical, online advice concerning some of the presented topics. This enhanced analysis allows a critical reading of the production landscape of popular science web videos on a global scale, helping describe the often understated connection between production and intention.

**Methodology**
**Selection of YouTube Channels**
For the selection process, we used the 'worldwide' list on the YouTube 'Science & Education' channel category site **[2]** and selected the most popular channels from 76 different countries (as of March 2015, the period of data collection). We compared the national channel lists with the global list and with recommendations from major science blogs to achieve a reliable selection of channels that include popular national channels, not only in English but also in Spanish, French, Portuguese, German, and Italian.

Firstly, we disabled cookies and cleaned cache memory data for searching. These factors interfere with the reliability of the findings due to search personalization settings and the effects of the so-called filter bubble (Pariser, 2011). Secondly, we loaded the 'worldwide' list on the YouTube 'Science & Education' channel category site. **[3]** From the end of 2012 to the time of data collection, this site worked with an algorithm that took not only views and subscriptions into account, but also user engagement. **[4]** This procedure allowed for the compilation of a global list of the one hundred most popular YouTube channels globally. Thirdly, we made a comparison by country. As of 18 March 2015, the time of data collection, it was possible to choose from among 76 countries. As a result, national and foreign science channels that were popular in the selected country were displayed. We compared these results with the global list of the most popular science video channels and include the most popular national channels in Spanish, French, Portuguese, and Italian.

The selection of video channels was supplemented by information that we retrieved from highly frequented science blogs, including Open Culture, Getting Smart, Make Use Of, MathsInsider among others. We identified 63 science blogs by means of Google searches using the following terms: '(best) youtube science channels', '(best) youtube educational channels', 'science blog youtube', 'recommend(ed) science channels', and their corresponding translations in German, French, Spanish, and Portuguese. 31 English, 15 Spanish, 13 German, and four Portuguese blogs were consulted. Expert recommendations on these blogs helped us to triangulate our observations and to choose the channels of seemingly greater impact, in terms of how often a science video channel was mentioned on the blogs consulted.

We excluded some video channels that did not fit our definition of "popular science web video" (see footnote 1). For the sake of consistency, the principal investigator reviewed all the web videos selected for inclusion. The resulting list includes 95 video channels. From each channel, we chose the most recent and the most popular video for analysis. Our study on the producers of popular science web videos is therefore based on a corpus of 190 web videos.

**Design of Coding Categories for Data Collection**
For the present analysis, we collected specific data about the most popular and the most recent video from each channel:

  a) Professionalism: video quality, audio quality, productivity, and profitability. We assessed the professionalism of the productions by crossing data on video quality with productivity and profitability. As for the latter, we inferred whether a video was produced for short-term profit or not, from the activation of advertising. **[5]**

  b) Type of producer: individuals or groups/organizations. With this data, we were able to tell if the video was the work of an individual or made by a group of two or more people, which we considered to be an 'organization'.

  c) Producer's gender and age profile: Number of producers (actors and presenters) by gender and by estimated age, to survey differences among them.

  d) Community building: we collected information on the position and type of recommended links for assessing the strategies for community building.

The age of the actors was estimated from facial and voice aging features. For age-assessment, we decided to reduce the distribution of population by age group to eight categories instead of the eighteen categories that we see in demographic pyramids. There are two reasons for this reduction in age categories. Firstly, the youngest age groups can be excluded since children under 13 are a special case due to legal restrictions and parental tutelage. Secondly, since we use a subjective method for assessing age, it is more reliable to allocate persons to ten instead to twenty-year age groups, which reduces the margin of error. Besides, we calculated a Cronbach's alpha to prove data consistency, obtaining the following satisfactory results: 0.7 for the age of female producers (see Figure 11), 0.8 for the age of male producers (see Figure 12), and 0.8 for the age of all producers (see Figure 10).

The participants in the analysis were trained to understand and correctly identify what we were looking for. The team was composed of one trainee (Friederike Shymura), two assistants (Thi Hoai Thuong and Klarissa Niedermeier) and two researchers (KlemensCzurda und Jesús Muñoz Morcillo). Reliability tests of 20 videos were conducted with an accuracy of more than 80% for each variable.

**Defining Professionalism:**
**Blurred Boundaries between Professional and User-Generated Content**
The difference between PGC and UGC is becoming less and less apparent due to the commodification of video production on the Internet. Since each producer is also a user, it is no longer productive to make a distinction between PGC and UGC. Instead, we opted to focus on the degree of technical expertise, production frequency and commodification, with a view to defining a new concept of professionalism, based on the present sample. Therefore, we recommend to replace the UGC category with non-PGC, if professionalism is understood as a combination of calculated measures for achieving popularity.

# Results

## Parameters for Professionalism: Analyzing Video Quality, Production Frequency, and Commodification in the Present Sample

Figure 1 shows the **quality of the assessed videos**. The results show that there are more productions which do not meet the 'High-Quality' (HQ) video criteria (57%) than those which do (43%); i.e., with at least a 720 pixels of vertical resolution and good sound quality. Focusing on the perception of audio quality, in Figure 2, we see that most of the videos have good (50%) or even very good (43%) audio quality. The high audio quality may seem surprising since most web videos are supposedly produced by "amateurs" (Keen, 2007: 5; Lovink, 2011: 9; for science web videos see Welbourne and Grant, 2015). However, we define amateurs in the YouTube context as those producers whose audiovisual know-how and storytelling skills are below the standards of average online media production skills. This can be assessed through the degree of montage complexity, the use of some professional methods (such as manual white balance, studio lights, or special effects), and the focus on good storytelling (Muñoz Morcillo et al. 2016). Within this new framework, some institutions could be considered amateurs, whereas some independent YouTube creators without institutional connection would not. Therefore, we also compared the video quality results with other data on production frequency and commodification.

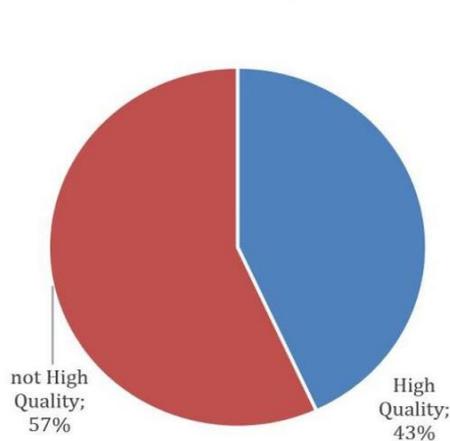

Figure 1. Video Quality According to Audio and Visual Quality (190 Videos)

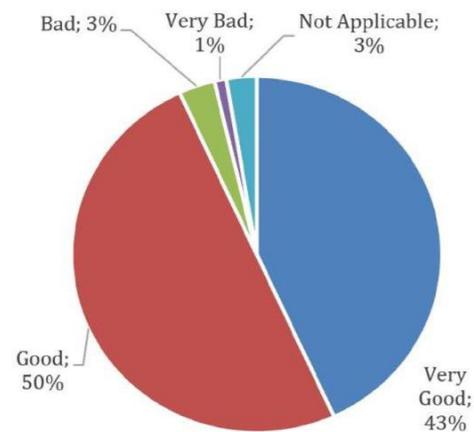

Figure 3. Video Production per Week (95 Channels)

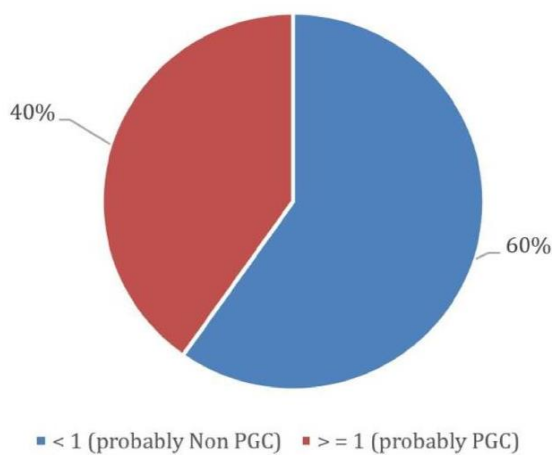

Figure 2. Audio Quality (190 Videos)

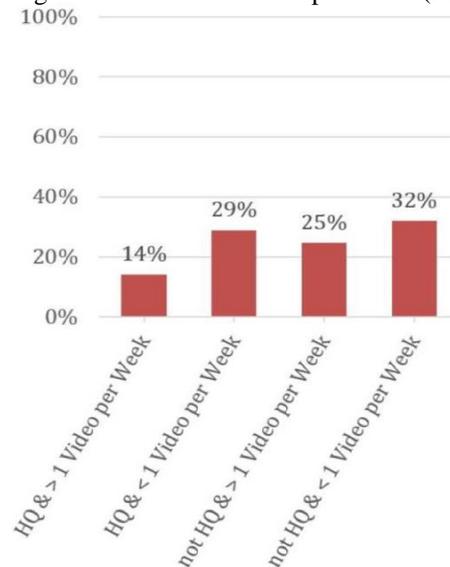

Figure 4. Relationship between Video Quality and Productivity (190 Videos)

The next index that we have considered is **the channels' production rate** (see Figure 3). When estimating the frequency of video production, we considered the production of more than one video per week as a possible starting point for nascent professionalism, or an intension to seriously pursue professionalism. In this case, we obtained similar results to those showing the difference between HQ and non-HQ videos, but with a slightly greater percentage of supposed non-professional producers (60%). Nevertheless, 40% of the analyzed channels have produced more than one video per week (Figure 3). However, regular production must correlate with other parameters such as good storytelling and community building measures, to contribute to a successful commodification strategy.

In the next step, we focused on the **relationship between audiovisual quality and production rate** (Figure 4). With regard to professionalism in the production of popular science web videos on YouTube, by crossing the data on audio and video quality (Figures 1 and 2) with the estimated frequency of video production per week (Figure 3), we can see that high-quality audiovisual videos are not always produced by channels with above-average video production (Figure 4). Just 14% of the sample are HQ productions, with an average production of more than one video per week. There are other factors involved in professionalism beyond audio and video quality, such as storytelling and dramatic devices (Muñoz Morcillo et al. 2016). This could also be the reason why non-HQ productions can sometimes be considered professional. Indeed, 25% of the videos from our sample belong to channels that produce more than one non-HQ video per week.

Regarding the 32% of the videos that are neither HQ productions nor belong to channels producing less than one video per week, we can consider them professional as long as their creators have other channels or receive so many views that they are not obliged to make one or more videos per week. This is the case with Derek Muller (Veritasium), Brady Haran (Sixty Symbols), or Destin Sandlin (Smarter Every Day). Therefore, if we assess the degree of professionalism, we would favor the idea of 'successful productivity' (in terms of popularity) at the expense of audiovisual quality.

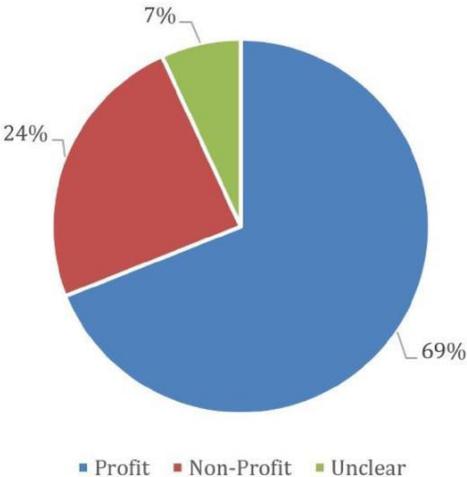

Figure 5. Profit versus non-Profit Producers (95 Channels)

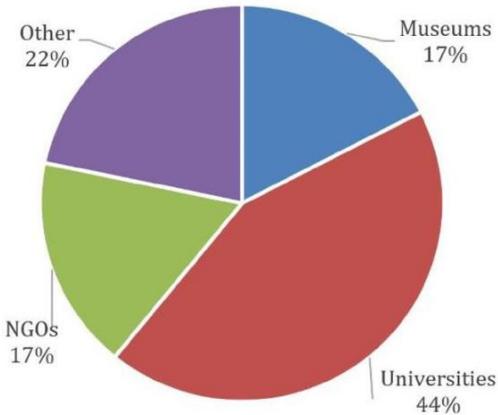

Figure 6. As Perceived Non-Profit Producers, i.e., without ads (46 channels)

Focusing on the channels that we have inferred as profit or non-profit from their perceived **advertising measures** (Figure 5) reveals that 69 % of all producing channels are profit-oriented. This means that even YouTube producers with underrated video quality or intermittent video production seem to monetize their content with advertising measures. Therefore, we cannot define a fixed professionalism index. It could be 40% of the analyzed videos, if we consider the video production rate as the preferred index of professionalism (see Figure 3), 43% if we prefer to consider the video quality (see Figure 1), or 69% if we choose profit-orientation (see Figure 5).

Although crossing these variables would give different qualities and levels of professionalism, the most prominent type of professionalism would be a video that brings together high audiovisual quality, regular productivity, and commodification measures. This is the case with asapSCIENCE, SciShow, SpanglerScienceTV, SickScience, or BozemanScience. There is another, more moderate but also effective type of professionalism which has regular productivity and commodification measures, but does not offer outstanding audiovisual quality. This is the case with channels such as Numberphile, Periodic Videos, or Unicoos. Videos with HQ and commodification measures but with an intermittent production rate can also be considered professional productions if the video functions as a showroom for attracting potential customers. An example of this would be kurzgesagt or Spacerip. Even productions without clear commodification strategies but with above-average video production such as TED Talks, TED-Ed, or Khan Academy (which does not even meet clear HQ standards) can be considered professional productions. In these cases, their commodification model works via donations and revenue from events and other offers on their respective platforms. Finally, videos that are not regularly produced, and without commodification measures, could be considered less professional in general terms, even if they were produced with HQ standards. This would be the case with Northwestern NewsCenter or Welt der Physik; though it would not apply to very successful channels that focus on storytelling, dramatic devices, and quality content, such as Periodic Table of Videos, Smarter Every Day, or Veritasium.

We identify producers who do not activate the YouTube monetizing tool as non-profit producers. The data illustrated in Figure 6 shows that 44% of the non-profit channels are university projects, 17% belong to Museums, and another 17% to NGOs (including organizations like TED). 22 % of the non-profit channels that do not monetize their videos, belong to institutions, for example research centers like the ESA (European Space Agency) or NASA, broadcasting companies, or television shows such as 'Abenteuer Wissen'. We could not clearly identify any channel among the non-profit producers run by an individual. On the other hand, 72% of the producers in our sample seem to be organizations (Figure 7), which we define as teams consisting of two or more people.

**Gender and Age of the Producers**

In order to determine the perceived gender-related information of the producers, we first documented the male and female presence in the analyzed videos (Figure 8). We found that in the most popular science web videos, visible producers such as presenters and actors were predominantly male. Just 24% of the producers were women in this sample, which corresponds to a gender gap of 26%.

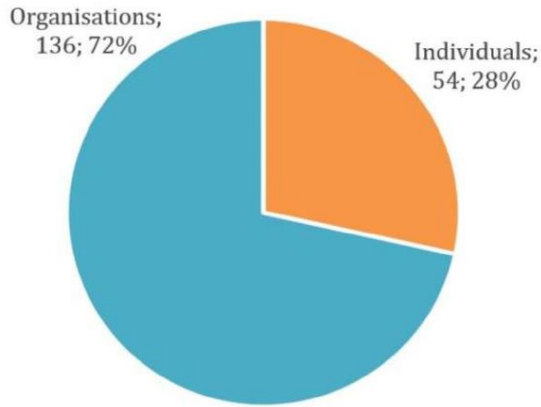 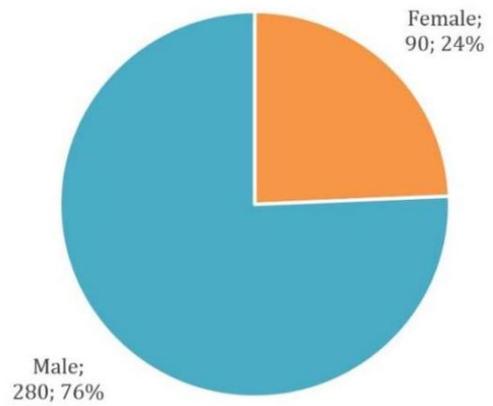

Figure 7. Producers' Profile (Individuals and Organizations, 190 Videos, Absolute Figures and Percentage)

Figure 8. Producers' Gender (370 Producers, Absolute Figures and Percentage)

Relating these results to the corresponding gender distribution in organizations and individual productions (see Figure 9), more women appear in videos produced by organizations (20%) than in individual productions (4%).

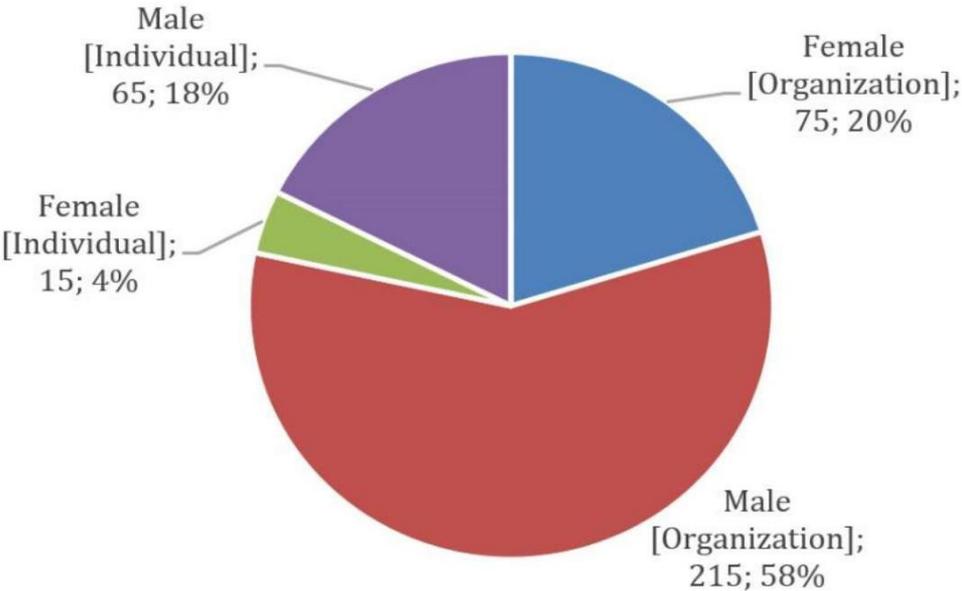

Figure 9. Producers' Gender According to Type of Production (370 Producers)
Absolute Figures and Percentage

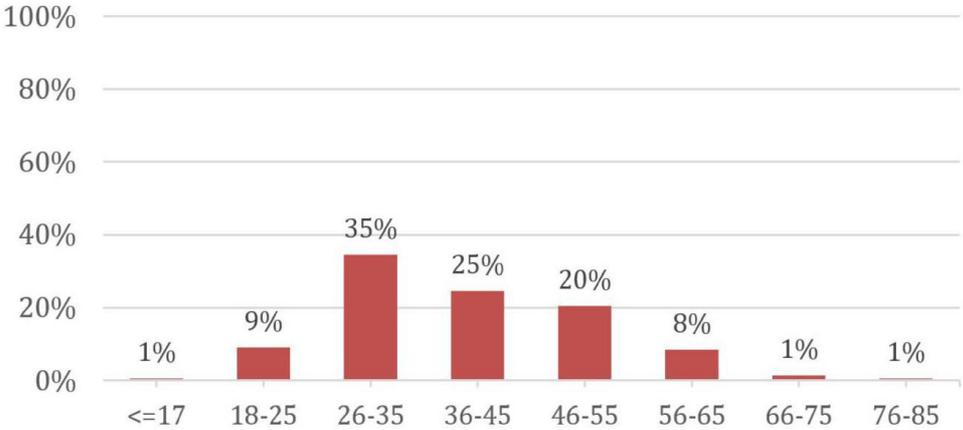

Figure 10. Average Age of All Producers
(142 Producers, i.e., Only Actors and Presenters/Producers)

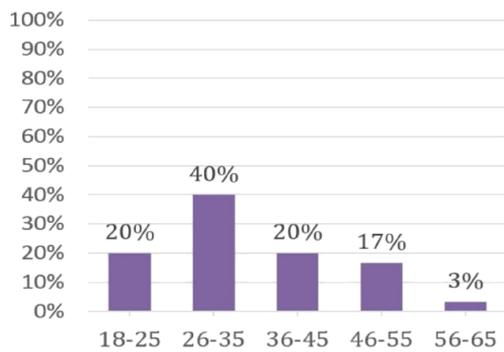 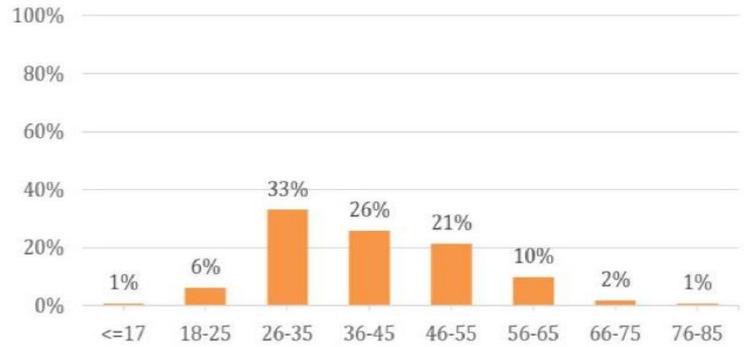

Figure 11. Average Age of Female Producers (30 Producers, i.e. Only Female Actors and Presenters/Producers)

Figure 12. Average Age of Male Producers (112 Producers, i.e. Only Male Actors and Presenters/Producers)

Regarding the age of the producers (Figure 10), the majority are between 18 and 45 years old, confirming the conventional wisdom that new technologies are for young people. Female producers partially fit this picture, since the majority of them are even younger, between 18 and 35 years old (Figure 11). Nevertheless, almost 20% of female producers are estimated to be between 36 and 45 years old, and 17% are between 46 and 55 years old (Figure 11). These two female age groups are slightly below the average in relative terms (Figure 10), which could also be a consequence of the small sample of female producers. We also observe a striking lack of female producers under the age of 18 and over 55.

In contrast, a higher number of male producers were recorded in older age groups (Figure 12). Nevertheless, most male producers are between 26 and 35 years old, followed by two age groups in similar proportions: 26% for 36-45 year-olds and 21% for 46-55 year-olds. The most noteworthy differences are the relatively small group of very young presenters, with just 6% of male producers between 18 and 25 years old; and the presence of other age groups that were absent among female producers, with 2% in the 66-75 age bracket and 1% in the 76-85 age bracket. This data indicates the existence of a significant gender gap in the production of popular science web videos in almost every age group, even when talking about relative values. If we look at the absolute figures, this gap is even more significant.

## Community Building

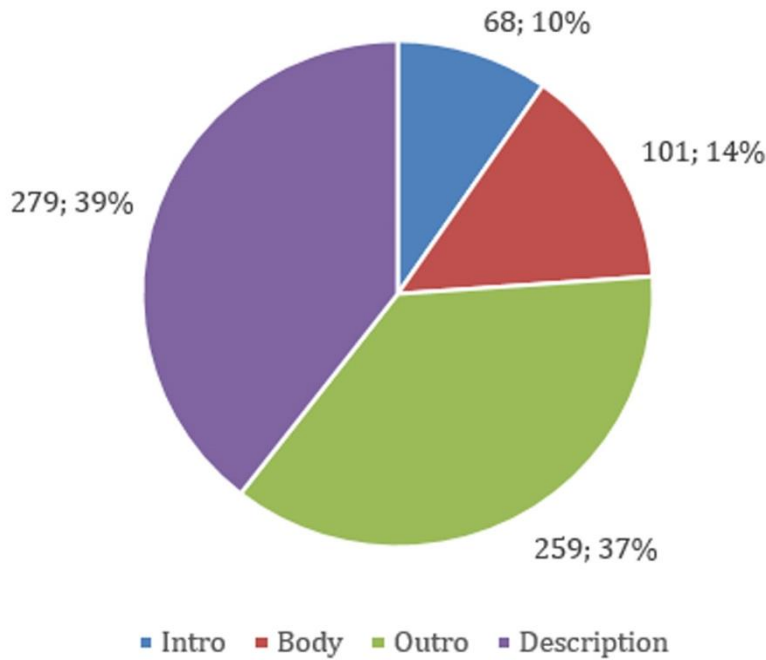

Figure 13. Number of Links per Position
(707 Links in 190 Videos, Absolute Figures and Percentage)

Let us now focus on the use of organic links. This is a crucial community building strategy which can also be considered as part of a profitability strategy, since community building is aimed at securing the future of a channel. This tendency is especially evident from producers' efforts to make recommended links visible. Figure 13 shows that most of these links, which include links to social media platforms, producers' web pages, or further videos from the same producer, can be found in the outro (37%) as well as in the video description (39%).

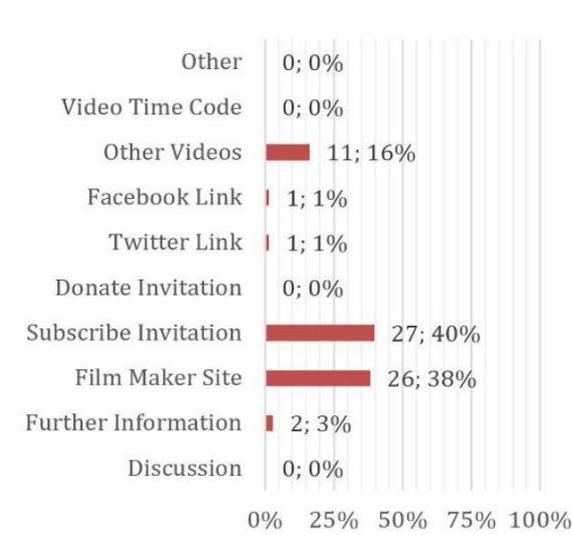

Figure 14. Type of Links in the Intro Sequence
(68 Links in 190 Videos, Absolute Figures and Percentage)

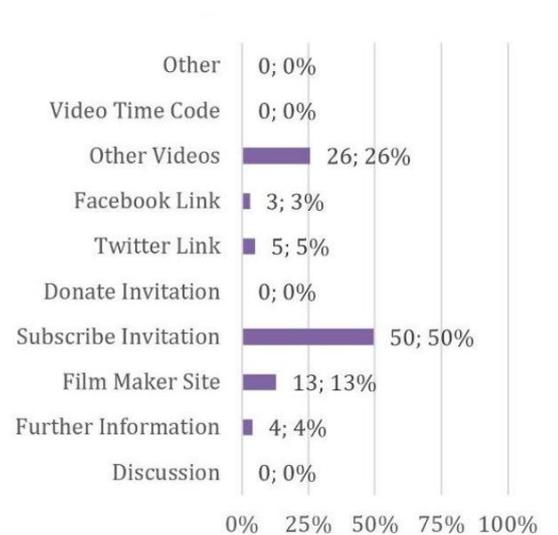

Figure 15. Type of Links in the body of the Video
(279 Links in 190 Videos, Absolute Figures and Percentage)

Nevertheless, we found invitations to subscribe, as well as links to the producers' web page in the intro sequence of many productions (Figure 14), although a higher number were recorded in the outro sequence (Figure 16).

Interestingly, the number of links and invitations to subscribe is even higher in the body, i.e., the main sequence of the film, than in the intro sequence (Figure 15). Recommendations and links could be found in the body of videos produced by very different channels such as SpaceRip, Imba Torben, The Slow Mo Guys, UNSW TV, It's Okay to be Smart, Smithsonian Channel, or Getty Museum. It is important to observe that most of these invitations to subscribe are YouTube annotations or watermarks that fade in at a certain point in the video. In some videos, we could observe this kind of subscribe-link annotation in the top left corner, in the last minutes of the body before the outro (e.g. Getty Museum). Whereas it mostly appears immediately after the intro, lasting until the end of the video as a watermark in the bottom right corner (e.g. Smithsonian Channel, It's Okay to Be Smart, The Slow Mo Guys, UNSW TV). They almost always appear in an appropriate position that barely interferes with the flow of the moving pictures, with many even adopting a branding function.

While other link categories, such as recommended videos, Facebook, Twitter, donate buttons, or third-party links have an intermittent or even a marginal appearance throughout the whole video structure, the invitation to subscribe seems to be a constant feature of the videos, with an enhanced presence in the outro sequence (see absolute figures in Figure 16).

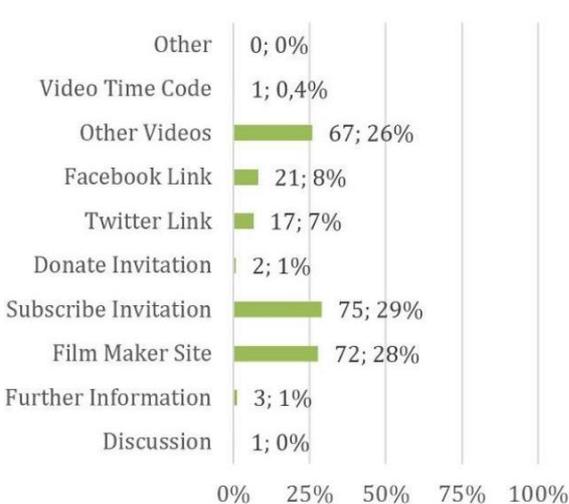

Figure 16. Type of Links in the Film's Outro (259 Links in 190 Videos, Absolute Figures and Percentage)

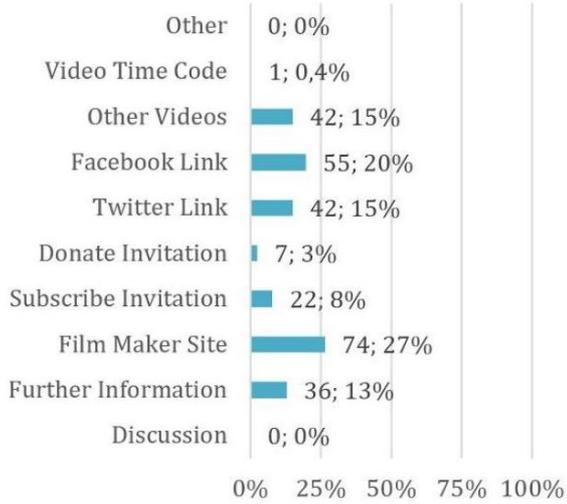

Figure 17. Type of Links in the Film's Description Area (279 Links in 190 Videos, Absolute Figures and Percentage)

The outro sequence is mostly used for recommending other videos and for references to the producer's web page. Here, we find 75 recommendations, more than in any of the other three positions that we analyzed.

In the description area (Figure 17), we could find examples of almost every kind of link, even "exotic" ones (at least within the context of our sample) such as time code links.

**Discussion**

*Professionalism no longer Makes a Distinction between UGC and PGC*

The first major topic of our survey was professionalism and its relation to the economic activities of YouTube producers. In our attempt to define a new concept of professionalism, we complement our assessment of a producer's audiovisual skills with other criteria, such as production rate and commodification strategies; all the while recognizing the importance of storytelling skills, as advocated by Muñoz Morcillo et al. (2016). Following this new definition, many institutions with quality resources for video production but poor storytelling skills could be defined as 'amateurs', i.e., users who are below average for online video literacy. Despite the elevated percentage of high quality audio and video in our sample, the degree of professionalism also depends on factors related to the success of the channel in terms of popularity.

In correlating the channel views and the number of subscriptions per 30 days, we have revealed some general insights into producers' strategies to enhance popularity and commodification. There is a significant number of channels producing less than one video per week with good or even very good audiovisual quality (29%, Figure 4). While this may seem confusing, a reasonable explanation for these results could be that producers are using their channels as a showcase for non-YouTube related work. Moreover, the commodification index, understood as the activation of advertising, paints an even more interesting picture. Almost 69% of all producers try to make a profit via advertising in their videos, but just 40% have what we could interpret as a serious professional strategy to regularly produce videos (i.e., almost one video per week).

It is more difficult to find a connection between technical professionalism and profit-oriented producers. Indeed, the next factor that we took into consideration for determining the perceived professionalism was the existence or absence of a commodification strategy, as evidenced by the activation or omission of advertising. The results show that even YouTube producers with inferior video quality or intermittent video production monetize their content with advertising measures. Therefore, we argue that there is a high level of professionalism among popular producers of science videos, while at the same time, a certain level of permeability between the categories of professional and non-professional production remains.

Additional qualitative data is needed to gain further insight into the platform-specific definition of professionalism. Nonetheless, we assume that there is a difference between some additional earnings and a clear profit. If earnings are to be understood as the main index of professionalism, then it could only become a substantial value when the producing channel can obtain a regular income for its services.

*Non-Profit Producers are Large Organizations*
The level of perceived commodification does not necessarily correlate with the level of professionalism. Most non-profit channels (i.e., without advertising) belong to universities, NGOs or research institutions. These channels have tax-based or fee-based business models, as opposed to the revenue-based models of individual producers on YouTube. Thus, we have not found even one individual video production which was defined as a non-profit activity.

We found out that all non-profit producers are large organizations such as universities, research centers, NGOs, or broadcasters. The reason for the absence of individual non-profit channels may be due to the limitation of our sample. Considering the large investment of time and money (Geipel, 2017) individual channel producers and small organizations may use whatever they can earn from advertising as a means to cover their costs. However, another interesting point is that most of the producers on YouTube are not individuals at all (i.e. one-person teams or YouTube creators acting as such), but rather organizations (i.e. teams of two or more people, see Figure 7). By assessing the credits, links and respective web pages, we found that 72% of the surveyed channels are run by organizations of two or more people. There are of course many individuals on YouTube producing and sharing their content for free, but just a small fraction of these are part of the most popular science web videos. Therefore, the popularity of YouTube science communicators is probably neither a haphazard phenomenon nor the result of purely altruistic activity, but rather the achievement of competitive, well organized YouTube "bestsellers."

*Success in Terms of Popularity is linked to Production Frequency*
We know that success is an elusive concept. Welbourne and Grant (2015) define success based on quantitative data such as views, subscribers, and others. Another factor for success could be earnings via advertising. In any case, the production rate seems to support both the attention-related and earnings-related definitions of success. However, for some YouTube producers, monetization does not seem to be as important as the popularity that can be reached by offering new and extraordinary content (Erviti and Stengel, 2016: 11) or the enjoyment of the producer's own passion to "edutain" their audience (Geipel, 2018). YouTube producer Brady Haran also underlines the importance of regular uploading and regular video production as one criterion for success on YouTube. This is consistent with our own observations on successful web video production (Erviti and Stengler, 2016: 9; see also Figure 3 and the corresponding comment).

*There is a Clear Gender Gap in Almost Every Age Group*
The second major topic of our paper was the analysis of age and gender distribution. The distribution of male and female producers is not at all symmetric. Female presenters of popular science web videos are highly underrepresented. This imbalance seems to be even more marked when examining the respective age groups, as well as their distribution in productions by individuals or organizations. Male YouTube producers have a much greater presence in general, outnumbering females in all but the 18-25 age group, where women and men are almost equally represented (6 women and 6,7 men). However, in this case, we must stress that there are only 30 female presenters in the whole sample, against 112 male presenters. Discussion among

YouTube creators about this gender issue suggests that offensive and sexist comments may be the primary reason for the low number of female science web video producers. Nevertheless, it seems that female presenters are slightly more visible in videos produced by organizations than by individuals. This is probably in line with many research institutions' gender equality policies, or because acting for an organization does not imply a high level of individual exposure, thus making it easier for female scientists to contribute to a web video production for the public.

We identified a considerable gender gap in the most popular science web videos since the presenters are predominantly male. Indeed, the gender gap of 26% in our sample is somewhat alarming given that the global Internet user gender gap was only 12% in 2016, according to statistics from the International Telecommunication Union (ITC, 2016: 3). This would suggest that the gender gap is greater in the production of popular science communication.

However, the question of gender stereotypes, where women present videos about beauty and lifestyle, and men talk about science and technology, is well-known among the YouTube community. One example is a recent discussion about the role and presence of women in tutorials, gaming videos, and other genres, which are predominantly controlled by (young) men (Meimberg 2016). Emily Graslie, presenter of the science related channel BrainScoop (2013), ascribed the reduced presence of female producers in popular science web videos to the virulent sexist comments they must deal with.

The absence of female producers under the age of 18 is probably not only relevant for science communication on YouTube, but for YouTube in general, since most of the platform's users are between 16 and 24 years old (Statista, 2016, Klicksafe, 2017). This may vary from country to country; in Germany for example, a statistical survey from Ipsos MediaCT states that more than half of the users are above 35 years old (Meedia, 2014).

We can summarize that the average age of male producers dominates the average age of all producers since the proportions barely change in the main diagram (cf. Figures 10 and 12). The most exciting thing we want to highlight is that there is no predominant age group for producers of popular science web videos on YouTube. The age groups of 26-35, 36-45, and even 46-55 are similarly represented; the most active among them being 26 to 35 year-olds (33%).

### *Community Building through Organic Links*
Recommending further links also seems to follow a profitability strategy. Most of the recommended links, including social media platforms, the producer's own web pages, or other videos, are placed mainly in the outro sequence, followed by the description, which is in line with recommendations from tutorials and YouTube themselves (cf. Dreier, 2015, from minute 27:00 on; YouTube Creator Academy, 2017b; Jenna Redfield Designs, 2017). This branding phenomenon connects 'organic links' with a strategy for community building. Hence we can find watermarks with invitations to subscribe in the body of many videos, occupying the space where traditional TV broadcasting companies would place their logos.

The intro sequence is decisive for transmitting credibility and curiosity in a very short period of time. Experienced YouTube producers try to avoid superfluous or obtrusive information. Placing an invitation to subscribe or a link in the very first seconds of a video could have a dissuasive effect on the user-viewer. Nevertheless, some of the most successful YouTubers include links at the beginning of their contributions. We can therefore assume that putting those links and recommendations in the intro sequence is a conscious decision made by science communicators, aware of the limited time they have to catch their audience's attention. This is in line with the observation that even professional producers with excellent audiovisual skills use the intro sequence for recommendations, invitations to subscribe or links to their own sites. One possible reason for the use of such information in the intro sequence may be the necessity of community building at all costs, in order to ensure the success of a channel. Another explanation could be the fact that the producers' web pages in the form of a logo with a website address, or invitations to subscribe using elaborate interfaces or watermarks, could also be part of the channel specific 'corporate identity'. Nevertheless, we cannot categorically dismiss the possibility that this practice may be no more than a typical beginner's mistake. However, since we are dealing exclusively with popular science web videos, it may be unlikely that the publication of links in the intro sequence is due to a lack of experience, since we can observe that very experienced and extremely popular channels such as TED-Ed, TED Talk or The Slow Mo Guys follow the same practice.

In our sample, placing recommendations and links in the outro sequence probably ensures community building. We argue that, from the viewer's perspective, the end of a video is often the beginning of a new search, facilitated (or even guided) by the producer's recommendation of related content. Given the above-average use of video recommendations, invitations to subscribe, and producer's web page in the outro, we maintain that the outro sequence has a unique, tactical potential for community building.
In a 2015 panel discussion on "Using YouTube for Original Content Distribution", Andy Stack, YouTube's former Head of Creator Technology, explained that collaboration between YouTube Creators is beneficial for everyone – YouTube, Creators and Advertisers – and that the reason for this may be the user's expectations after having watched interesting content. Indeed, at the end of a video, an interested viewer often follows the 'organic links' provided (such as invitations to subscribe and related videos) thereby continuing the 'viewing conversation' (cf. Dreier, 2015, from minute 27:00 on). Similar explanations could be found in several tutorials from YouTube's Creator Academy (YouTube Creator Academy, 2017b) as well as on independent blogs which provide recommendations for success on YouTube (Jenna Redfield Designs, 2017).
We also argue that the location of invitations to subscribe and other information even in the main part of the video is evidence that the producers are following an economic strategy. This is in line with the assumption that producers must adapt to YouTube's platform policies to gain success (Geipel, 2018).

Finally, the description area is an essential part of YouTube's video page structure. However, users only see a summarized video description when choosing from a list of search results. Here, a preview thumbnail appears alongside additional information about the video, including the title, the user's name, the first lines of the description, as well as some basic statistics like the number of views or the date of publication. We, therefore, assume that the details of the description are not the first thing that an average user would notice, at least not before having watched the whole video or parts of it. This is probably the reason why we do not perceive any clear strategy for monetizing information in the description area, despite the fact that the more keywords it contains, the better the SEO (search engine optimization, cf. Pinsky, 2014).

On the other hand, the citation of sources and further reading is probably a good indicator of a channel's trustworthiness but it may also have some branding function (cf. Pinsky, 2014). Furthermore, YouTube strictly controls the links within its videos, only allowing those which refer to content within YouTube itself; whereas the description area also permits links to external sites.

**Conclusions**

Our analysis of 190 popular science web videos provides a general picture of some fundamental characteristics of the production of popular web videos on YouTube, focusing on the topics of professionalism, producer's gender and age profile, and community building. We have discussed our findings with up to date qualitative research as well as information from particular blogs or official YouTube web pages.

As a novelty, we have departed from the old distinction between User Generated Content (UGC) and Professional Generated Content (PGC) in favor of a more accurate distinction between professional and non-professional productions; avoiding the heretofore misleading tendency to equate the term 'user', practically a synonym for 'amateur', with 'non-professional', when analyzing web video production.

Our discussion on professionalism has shown that the old distinction between UGC and PGC no longer seems to fit. In disconnecting the term professionalism from the definition of professionals versus users, we now use objective criteria such as audiovisual quality, production rate, and commodification for an appropriate definition. However, crossing some of these results has revealed that the limits of professionalism are more complicated to define than we initially thought. Some very successful channels that we usually identified as professional have neither high quality (HQ) video nor a regular production of one video per week. Either their creators have other channels and are very successful with lower production rates than one video per week, or they have other business models that allow them to avoid Google ads or regular weekly production. Other parameters should also be considered, especially the use of dramatic devices, which is a good indicator of how an individual YouTube creator, or an organization, understands the art of filmmaking. However, our discussion on professionalism has shown that professionals need to be successful in order to earn money whatever their business model. Therefore, linking success to production frequency and dramatic, rather than audiovisual, quality could be a suitable indicator for identifying professional science web video producers. Since our sample was based on popular science web videos, the majority of them are successful productions in terms of popularity; therefore for any future analysis on the distinction between professional and non-professional science web videos we would need a broader sample of videos.

We have also noticed that the non-profit producers are large organizations, such as universities, research centers, and NGOs. However, these have other business models and revenue streams beyond YouTube. Universities need to be authentic and credible in order to attract students and funding, while NGOs such as TED Talks organize regular events in different countries.

One of the most sobering results is that there is a gender gap in almost every age group in our sample. Male producers dominate the popular science web video scene, since female creators are a minority in every age group we have analyzed. In our discussion on the gender gap, we have noticed that very successful female YouTubers such as Emily Graslie (BrainScoop) allude to online sexual harassment and abusive language in YouTube comments as the primary cause for this significant gender gap.

Irrespective of gender issues, one of the most important aspects seems to be the creation of exciting content capable of supporting community building measures, mainly through organic links which are mostly concentrated in the outro. The location of invitations to subscribe in every part of the video is evidence that also science communicators follow an intense economic strategy adapting to YouTube's game rules. Even the description area is mostly designed in order to generate trustworthiness and, therefore, to attract new followers by offering additional information.

Despite the fact that many YouTube creators may honestly pursue some societal and educational ideals, it seems that even the production of popular science web videos is mutating into a new form of traditional mass media, with the irruption of small (or not so small) vices, and transgressions. Undoubtedly, the most promising feature is the fact that popular science web videos are now part of our "being in the world" in the words of Merleau Ponty, that is to say, they have the potential to enhance our consciousness and sharpen our perception for good. The recommendation culture, the online discussions, the ubiquity and diversity of topics, and the blurred boundaries between users and producers are intriguing and challenging aspects of our understanding and experience of web videos. But ultimately, the quality of the societal contributions of this kind of media depends on the intentions of its producers.

**Notes**

**[1]** We define popular science web video as an online video focusing on the communication of scientific content to a broad audience.

**[2]** https://www.youtube.com/channels/science_education (accessed on 18 March 2015, since depreciated by YouTube). At the time of the data collection, this site displayed the most popular science channels worldwide and per country.

**[3]** https://www.youtube.com/channels/science_education (visited on 18 March 2015)

**[4]** Cf. http://youtubecreator.blogspot.de/2012/10/youtube-search-now-optimized-for-time.html (accessed on 8 November 2018).

**[5]** Some institutions also pursue long-term financial profit without using advertising. There are many ways of commodification via video production; we have chosen the advertising activation as an indicator, for the sake of simplicity and methodological consistency.

**Appendix**
**Table 1: List of 190 analyzed videos from 95 channels**

| Nr. | Channel | Video title |
|---|---|---|
| 1 | asapSCIENCE | Which Came First – The Chicken or the Egg? |
| 2 | asapSCIENCE | The Scientific Power of Teamwork |
| 3 | Khan Academy | Basic Addition |
| 4 | Khan Academy | Subtracting complex numbers |
| 5 | MinutePhysics | Immovable Object vs. Unstoppable Force – Which wins? |
| 6 | Minute Physics | How modern Light Bulbs Work |
| 7 | NurdRage | Coke Can in Liquid Nitrogen |
| 8 | NurdRage | Make an Iron Heart |
| 9 | Scientific American Space Lab | How to Enter the Space Lab Competition |
| 10 | Scientific American Space Lab | Behind the Scenes |
| 11 | SciShow | The Truth About Gingers |
| 12 | SciShow | Trouble in Bed: When Sleep Turns Against Us |
| 13 | SpanglerScienceTV | Magic Sand – Sand that is Always Dry! |
| 14 | SpanglerScienceTV | Lava Lamp – Cool Science Experiment |
| 15 | Sixty Symbols | Putting your Hand on the LHC |
| 16 | Sixty Symbols | The Sound of Atoms Bonding |
| 17 | Smarter Every Day | How Houdini Dies (in Slow Motion) |
| 18 | Smarter Every Day | Cold Hard Science. The Physics of Skating on Ice |
| 19 | SpaceRip | Earth 100 Million Years from Now |
| 20 | SpaceRip | Water Planet |
| 21 | TED-Ed | Questions no one knows the answer to |
| 22 | TED-Ed | Why do we cry? The three types of tears |
| 23 | TED Talks | Tony Robbins: Why we do what we do |
| 24 | TED Talks | Mary Lou Jepsen: Could future devices read images from our brain? |
| 25 | Veritasium | World's Roundest Object |
| 26 | Veritasium | Can you solve this? |
| 27 | Vi Hart | Hexaflexagons |
| 28 | Vi Hart | Cookie Shapes |
| 29 | SickScience | Dry Ice Boo Bubbles |
| 30 | SickScience | Power of Bleach |
| 31 | Acchiappamente | 2x05 – Disney e Coca-cola ti controllano? [Messaggi subliminali – Psicologia] |
| 32 | Acchiappamente | #Acchiappamente – Stress buono o cattivo? |
| 33 | Alberto Lori | Libertà di cambiare (psicologia quantistica) |
| 34 | Alberto Lori | Il pensiero focalizzato (psico quantistica) |

| 35 | ANUchannel | Richard Dawkins and Lawrence Krauss: Something from Nothing |
|---|---|---|
| 36 | ANUchannel | A Conversation with Andrew Macintyre |
| 37 | BozemanScience | A Tour of the Cell |
| 38 | BozemanScience | The Brain |
| 39 | Canal Educatif à la demande (CED) | Simulation d'entretien de recrutement |
| 40 | Canal Educatif à la demande (CED) | L'art en Question 08: Carpaccio – Le Jeune Chevalier |
| 41 | Deep Sky Videos | Messier Objects |
| 42 | Deep Sky Videos | Inside an Opening Telescope |
| 43 | La Educación Prohibida | LEP – Archivos Abiertos #01 – Carlos González |
| 44 | La Educación Prohibida | LEP – Archivos Abiertos #11 – Antonio Solórzano |
| 45 | GeroMovie | DNA-Replication Biologie |
| 46 | GeroMovie | Winkelarten |
| 47 | ImbaTorben | Todesmilch Titten |
| 48 | ImbaTorben | Die 5 A's – so bekommst du du jede Frau |
| 49 | matemarika86 | Non funziona la funzione!!! – Studio di funzioni e dominio |
| 50 | matemarika86 | VIDEO INTERATTIVO Caccia al tesoro: Alla ricerca della X perduta – Equazioni di primo grado intere |
| 51 | Mental Floss | 50 Common Misconceptions |
| 52 | Mental Floss | 27 Amazing Facts about Comics |
| 53 | NASA JPL Videos | Mars Science Laboratory Curiosity Rover Animation |
| 54 | NASA JPL Videos | What's Up for March 2014? |
| 55 | The Slow Mo Guys | Giant 6ft Water Balloon |
| 56 | The Slow Mo Guys | Airbag Deploying in Slow Mo |
| 57 | NewScientist | Spray-on Clothing |
| 58 | NewScientist | Cyborg Drummer creates Unique Beats |
| 59 | Nucleus Medical | Birth: McRoberts Maneuver |
| 60 | Nucleus Medical | Nucleus Custom Medical Animation Process |
| 61 | PBS IdeaChannel | Are Bronies Changing the Definition of Masculinity? |
| 62 | PBS IdeaChannel | Does Twitch Plays Pokemon Give You Hope for Humanity? |
| 63 | RMIT University | How hydrogen engines work |
| 64 | RMIT University | Australia-India Research Centre für Automation Software Engenieering |
| 65 | Scientific American | Your Brain in Love and Lust |
| 66 | Scientific American | Is Our Universe a Hologram? |
| 67 | Storm | Amazing Upward Lightning! |
| 68 | Storm | Extreme Dust Storm Takes Over Phoenix, Arizona 2011 |
| 69 | foodskey | Being mean to broccoli |
| 70 | foodskey | Phytoplasmas in Plants |

| 71 | Unicoos | Matriz inversa, traspuesta y adjunta 2ºBACHI unicoos matemáticas |
| 72 | Unicoos | BILLION = BILLON?? unicoos nosvemosenclase Facebook compra Whatsapp |
| 73 | Unisciel | Faire implose une canette |
| 74 | Unisciel | Unisciel Select: numero 55 |
| 75 | Universcience | FIV mode d'emploi |
| 76 | Universcience | Herbier #7 – on a une belle série de citrons |
| 77 | UNSW TV | How to survive beach rip current |
| 78 | UNSW TV | Why winds explain earth's surface warming slowdown |
| 79 | CrashCourse | The Agricultural Revolution |
| 80 | CrashCourse | Fate, Family, and Oedipus Rex |
| 81 | Computerphile | The Problem with Time & Timezones |
| 82 | Computerphile | EXTRA BITS – Installing Ubuntu Permanently |
| 83 | MinuteEarth | Where Did Earth's Water Come From? |
| 84 | MinuteEarth | Are any Animals Truly Monogamous? |
| 85 | Numberphile | Why do YouTube views freeze at 301? |
| 86 | Numberphile | Brussels Sprouts |
| 87 | PeriodicVideos | Gold Bullion Vault – Periodic Table of Videos |
| 88 | PeriodicVideos | The world's greatest autograph book |
| 89 | Vsauce | What if Everyone JUMPED at once? |
| 90 | Vsauce | What is the Resolution of the Eye? |
| 91 | The SpanglerEffect | Getting Ready for Guiness World Record Season 01 Ep.01 |
| 92 | The SpanglerEffect | Flying Toilet Paper Season 02 Ep.19 |
| 93 | CGPGrey | The Difference between the UK, GB and England |
| 94 | CGPGrey | The Law You Won't Be Told |
| 95 | Vlogbrothers | Giraffe Love and Other Questions ANSWERED |
| 96 | Vlogbrothers | Is the American Dream Real? |
| 97 | Quantum Fracture | ¿Qué es la Ciencia? |
| 98 | Quantum Fracture | Uno de los Principios Más Importantes del Universo |
| 99 | MinutoDeFísica | Errores comunes en física |
| 100 | MinutoDeFísica | E=mc² está Incompleta |
| 101 | Ever Salazar | Calculando Areas |
| 102 | Ever Salazar | (not so) Cold Fun: Qué hacer cuando no está tan frio afuera |
| 103 | The Spirit Science | Spirit Science 1 ~ Thoughts |
| 104 | The Spirit Science | Spirit Science 22 (part 4) ~ Source Energy |
| 105 | ScienceBob | Science Bob's Crazy Foam Experiment |
| 106 | ScienceBob | Exploding Pumpkins on Jimmy Kimmel Live |
| 107 | ouLEarn | Shakespeare: Original Pronunciation |
| 108 | ouLEarn | Maryam Bibi – Unlikely Leaders (2/5) |
| 109 | Euronews Knowledge | World's smallest atomic clock |
| 110 | Euronews Knowledge | Can Earthquakes Bring Life? Do You Know? |
| 111 | Naked Scientists | How does DNA fingerprinting work? Naked Science Scrapbook |
| 112 | Naked Scientists | Main Alu II re-uploaded |
| 113 | Educatina | Síntesis de proteínas – Biología – Educatina |
| 114 | Educatina | Patrones de medición – Física – Educatina |
| 115 | FavScientist | Gregor Mendel – My Favourite Scientest |
| 116 | FavScientist | Ignaz Semmelweis – My Favourite Scientist |
| 117 | Depfisicayquimica | Agua que no cae / The water doesn't fall down |
| 118 | Depfisicayquimica | Cubo que no se derrama II |
| 119 | Brusspup | Amazing Anamorphic Illusions |
| 120 | Brusspup | Cool Rolling Illusion Toy! How to |
| 121 | Jörn Loviscach | 22.6.1 Stetigkeit |
| 122 | Jörn Loviscach | P3 Datumsdifferenz in Tagen mit Embedded Controller |
| 123 | ChemExperimentalist | How to make sulfuric acid |
| 124 | ChemExperimentalist | Make Calcium Hydroxide – Ca(OH)$^2$ from Plaster of Paris |
| 125 | Abenteuer Wissen | Star Trek: Wie funktioniert Impuls- und Warpantrieb |
| 126 | Abenteuer Wissen | Timescapes: Learning to Fly – die Welt im Zeitraffer |
| 127 | Welt der Wissenschaft | Animaterie und Relativität |
| 128 | Welt der Wissenschaft | Wie einzigartig ist der Mensch? |
| 129 | Welt der Physik | Was ist ein schwarzes Loch? |
| 130 | Welt der Physik | Monsterwellen im Labor |
| 131 | WissensMagazin | Der tiefste Blick ins All |
| 132 | WissensMagazin | E=mc$^2$ – Die Äquivalenz von Masse und Energie |
| 133 | TheSimpleMaths | Exponentialfunktion und Logarithmus |
| 134 | TheSimpleMaths | Gebrochenrationale Funktionen – Nullstellen, Definitionsbereich… |
| 135 | Fisica Total | Física Total – Aula 07 – vetor – Vetores e operações vetoriais |
| 136 | Fisica Total | ENEM em AÇÃO – Física #01 (principais habilidades cobradas na prova de…) |
| 137 | Canal Ciência e Ficção | Jurassic Park parte 1/2 – É possível clonar dinossauro? – Ciência e Ficção |
| 138 | Canal Ciência e Ficção | Star Wars Episódio VII – O Que Esperar? – Ciência e Ficção |
| 139 | Manual do Mundo | Congele água em 1 seg – o segredo |
| 140 | Manual do Mundo | Revelação da Mágica dos ladrões de galinha (mágica fácil revelada) |
| 141 | Quirkology | 10 More Amazing Science Stunts (3) |
| 142 | Quirkology | The Tube of Mystery |
| 143 | NASA eClips | Real World: Space Shuttle Thermal Protection System |
| 144 | NASA eClips | Real World: Comet Quest |
| 145 | It's Okay To Be Smart | The Science of Snowflakes |

| 182 | TU Muenchen | TUM Ambassador Professor Patrick Dewilde about the importance of networks for scientific thinking |
| 183 | European Space Agency, ESA | First-ever live 3D video stream from space |
| 184 | European Space Agency, ESA | Sentinel-1A rides into space on a Soyuz |
| 185 | Kurzgesagt | Fracking explained: opportunity or danger |
| 186 | Kurzgesagt | Engineering & Curiosity |
| 187 | Hybrid Librarian | World's 10 Most Mysterious Pictures Ever Taken |
| 188 | Hybrid Librarian | Earth's 10 Most Important Events in History |
| 189 | Cambridge University | Memories of old awake |
| 190 | Cambridge University | Putting our House in Order: William Kent's Designs for the Houses of Parliament |